# Calculating Magnetic Properties of Two-Dimensional Materials for Memory Applications

Md Rakibul Karim Akanda, D'Mytri Wiggs, and Nathaniel Zepeda

Department of Engineering Technology, Savannah State University—Savannah, GA 31404, United States of America

**Abstract**

Calculating magnetic properties of two-dimensional materials is crucial for implementing memory devices (like USB drive, RAM, hard disk drive of computers) having reduced size. Two dimensional materials can be implemented as a thin film which can reduce the size of memory devices. These materials as well as devices made with magnetic two-dimensional materials are of current research interest in industry and academia. From the materials project database, crystal structure file of 30 two dimensional materials have been downloaded to calculate their magnetic properties. BURAI Quantum espresso software has been used to extract magnetic properties of 30 two dimensional (2D) materials. These 30 materials have magnetic Fe, Ni, Co, Mn, and Cr atoms in their molecular structure. Magnetic materials play a key and vital role in today's modern-day technology. There are five different types of magnetic materials. The classification magnetic materials are diamagnetism, paramagnetic, ferromagnetism, ferrimagnetism, and anti-ferromagnetism. Total energy of different magnetic configurations has been calculated to find the most stable magnetic configurations of these materials.

## 1. Introduction

BURAI Quantum ESPRESSO is a free first-principles electronic-structure calculator and materials modeler which is used to calculate different properties of the materials. ESPRESSO is an acronym for Open-Source Package for Research in Electronic Structure, Simulation, and Optimization. The software is based on density-functional theory (DFT), and is used to calculate total energy, and magnetic moment.

Two dimensional materials are first extracted from materials database. Then the materials were analyzed to find its magnetic properties. Materials have been identified as ferromagnetic or antiferromagnetic materials from the calculated lowest total energy which is also called ground or more stable state. Antiferromagnetic materials can store a larger amount of data compared to ferromagnetic materials. Magnetic moment is also calculated which shows the magnetic strength in a material. Extracted parameters and properties can be used to simulate new memory devices and smaller chips.

The name of the software used is called Burai 1.3 expresso, and it is excellent. This software will allow you to run different calculation methods such as SCF, band, and geometric and each of these will let you run different calculations and they all serve its own purpose. Quantum espresso is a suite for first-principles electronic-structure calculations and materials modeling, distributed for free and as free software under the GNU General Public License. It is based on density-functional theory, plane wave basis sets, and pseudopotentials. When it comes to running calculations for the software of Burai 1.3 the pseudopotentials are everything. The pseudopotentials are like your directions to a hard piece to put together or a label that's on the outside of a food box. When running calculations without the pseudopotentials the calculation will not run at all. This software is used to run calculations such as band structure, band, and SCF and find out if different objects and materials are magnetic and if they are magnetic are they antiferromagnetic or just ferromagnetic. Through using this you can also tell the band gap of the material and it is the gap in the different bands of the material if the material has multiple bands. If it has multiple bands and they are very spacious then of course the band gap would be high but if they are not to far apart then the band cap would be closer together and band gap would not be as high. At the end of the DFT calculations you are going to calculate all the ground state properties. Two electrons can interact with electronic charge, and they also can by their spin and mutual repulsions and attractions can be casualty for this. The main equations for calculating this is the Kohn-Shom equation. When you have a DFT package that you are trying to calculate give an input of atomic number position to make sure it is correct, then use any software to extract positions in a suitable format. A magnetic material in my personal way of viewing it is any material that has a magnetic field in it whether it is on the periodic table or not and if it is magnetic, it can be antiferromagnetic or it can just be regularly ferromagnetic. The magnetic field on a magnetic object is invisible but it holds so many responsibilities when it comes to the magnetic. The most notable property of a magnetic is a force that pulls on other ferromagnetic materials, such as iron, steel, nickel, cobalt, and attracts or repels another magnetic. These magnetic materials vary so wide in range to the point it may be difficult to pinpoint. They vary on a scale from ten to one hundred newton meters with a defined structure for a particular application. Magnetic materials play a key and vital role in today's modern-day technology. Without the use of magnetic materials in the world a lot of the resources and everyday life things we use would be harder to make or would not hold as much value as they would without the magnetic materials in them. For example, when you look at motors, transformers, and generators we all use these things in our everyday life and the key component of all these examples are the magnetic materials that are inside of them. The magnetic materials have been around for three thousand fiver hundred years, and it was first discovered in Magnesia. It has been around almost as long as the compass which has been around for four thousand five hundred years. The nineteen eighty's saw continued improvements in magnetic alloys, rapidly accelerating in thin films and surfaces, and the magnetic recording developments those activities supported, as well as the emergence of permanent magnets. The nineteen nineties were special because that's when they had done research and reached the emergence of one of the nicest magnetic materials which was the nanocrystalline magnetic material. This material is used for effective antimicrobial for treating wounds especially burns and chronic wounds. This also reduces the inflammatory processes and promotes would healing and is less toxic than other forms of silver dressings due to the prolonged release of silver to the wound.

Quantum ESPRESSO is a free first-principles electronic-structure calculator and materials modeler. ESPRESSO is an acronym for (Open-Source Package for Research in Electronic Structure, Simulation, and Optimization). The software is based on density-functional theory, plane wave basis sets, and pseudopotentials (both norm-conserving and ultrasoft). Quantum ESPRESSO was used to calculate and graph operations that are common in SCF (Self-Consistent Fields) and DFT (Density Functional Theory).

The conduction band is a delocalized band of energy partly filled with electrons in a crystalline solid. These electrons have great mobility and are responsible for electrical conductivity. Valence bands are the bands of electron orbitals that electrons can jump out of, moving into the conduction band when excited. The valence band is simply the outermost electron orbital of an atom of any specific material that electrons occupy. Band gaps are the minimum amount of energy required for an electron to break free of its bound state. When the band gap energy is met, the electron is excited into a free state, and can therefore participate in conduction.

Solid-state physics describes the electronic band structure by the range of energy levels that electrons may have within it, as well as the ranges of energies not found (band gap). They are typically depicted on a 2D graph by graphing the energies of crystal orbitals in a crystalline material. Conductors (metals) have no band gaps between the valence and conduction band, this means that electrons can move freely between valence and conduction bands. Insulators have large gaps between valence bands and conduction bands, this means that the conduction band is empty, and the material cannot conduct electricity. Semiconductors have small gaps between the valence band and conduction band. This allows for certain electrons (given sufficient energy) to pass through the valence and conduction band.

Density of states describes the proportion of states that are occupied by the system at each energy. It is represented mathematically through a probability density function and averages the space and time domains that its various states occupied by the system. Density of states of matter is also continuous.

Density-functional theory is the computational quantum mechanical modelling method used in physics, chemistry, and material science. DFT is used to investigate the electronic structure of many-bodied systems, mainly molecules, atoms, and their condensed matter. What is crystal structure in real space and k-points in reciprocal space. DFT calculations allow the equilibrium particle density and prediction of thermodynamic properties and the behavior of many-bodied systems on the basis model of interactions between particles. Spatially dependent density is used to determine the local structure and composition of materials.

A crystal structure is defined as the repeating arrangement of atoms (molecules or ions) throughout a crystal. Structure refers to the internal arrangement of particles and not the external appearance of the crystal. Crystal structures can be described in several ways. When described it usually refers to the size and shape of a unit cell and the positions of the atoms within the cell. However, insufficient information can cause a poor understanding of the true structure within three dimensions. There are several factors to consider such as the atoms unit cells arrangement, contact, and distance are used to create a full understanding. Solid state theory defines “k-space” refers to reciprocal space. However, in electronic structure theory k-points serve a more specific purpose. They are used as sampling points in the first Brillouin zone of material. This typically refers to the specific region of reciprocal space that is closest to the origin (0,0,0).

## 2. Procedure

Magnetic materials applications are the creation and distribution of electricity and are also in appliances that run using electricity. They are used for like storage of data on audio and video tape as well as on the computer desktop. There are five different types of magnetic materials. The magnetic materials are diamagnetism, paramagnetic, ferromagnetism, ferrimagnetism, antiferromagnetic. Ferromagnetic material are materials that exhibit a spontaneous net magnetization at the atomic level, even in the absence of an external magnetic field and some examples of ferromagnetic materials are iron, cobalt, and nickel. Ferrimagnetic materials are the populations of atoms with opposing magnetic moments. For ferrimagnetic materials, these moments are unequal in magnitude, so a spontaneous magnetization remains. Ferrimagnetic material do not have a plethora of examples, but the best example of this material would be magnetite. Antiferromagnetic is a magnetism in solids such as manganese oxide in which adjacent ions that behave as tiny magnetic fields spontaneously align themselves at relatively low temperatures into opposite, or antiparallel, arrangements throughout the material so that it exhibits almost no gross external magnetism. Examples of antiferromagnetic materials Hematite, chromium, alloys of iron manganese and oxides of nickel. Crystal structure is defined as the repeating arrangement of atoms throughout a crystal. Structure refers to the internal arrangements of particles and not the external appearance of the crystal. The simple cubic shell is just a cube with an atom on each corner. This also belongs to the space group two hundred and twenty-one. Face-centered cubic is an atom arrangement found in nature. This cell structure consists of atoms arranged in a cube where each corner of the cube has a fraction of an atom with six additional full atoms positioned at the center of each cube face. A body centered cube just like the face-centered cube is an atom that is found in nature. Its makeup is a cell structure that consist of atoms arranged in a cube where each corner of the cube shares an atom and with one atom positioned at the center this is how the body centered cubic works so well. When doing a spin polarized calculation first you will set the parameters on however you want to, or you can leave the parameters in the default, and you can use any pseudopotential when trying to calculate it. You will set the starting magnetizations and then once everything is set you will click on the run button and then save it and run it as a SCF calculation and then once the file is done running you will go to the output file and look at the total magnetization and watch how the graph is converging and it will show you the magnetization.

## 3. Results

| | | |
|---|---|---|
| 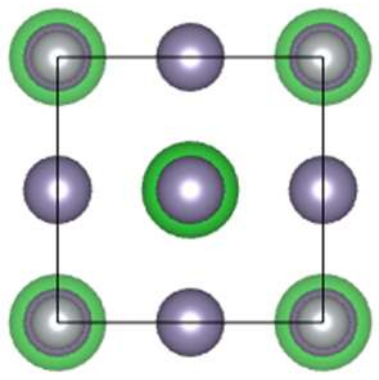 $BaNiSn_3$ | 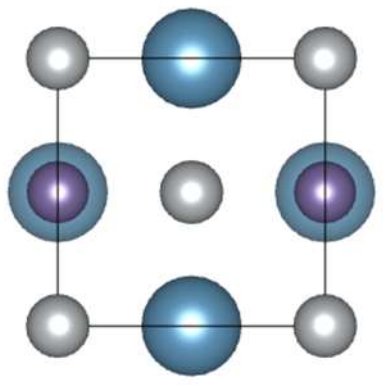 CaNiGe | 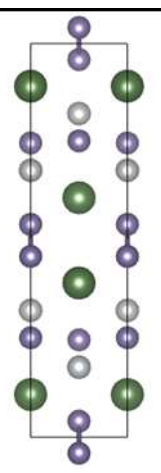 $ErNiGe_2$ |
| 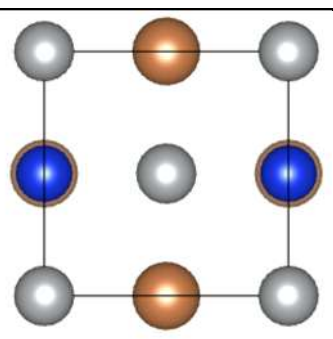 SiNiSb | 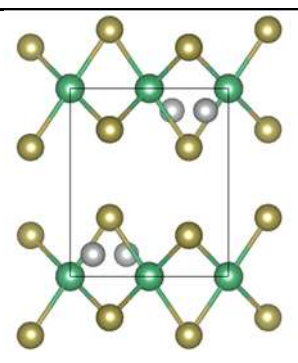 $NbNiTe_2$ | 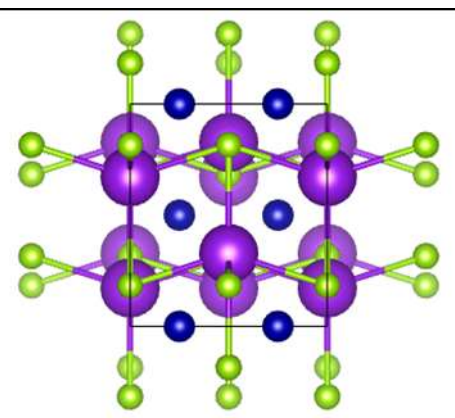 $K_2CoSe_2$ |
| 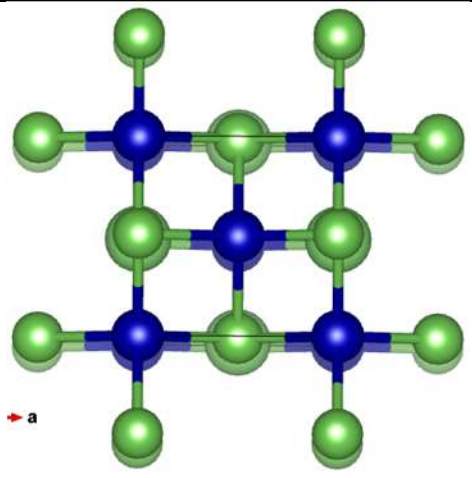 LiCoAs | 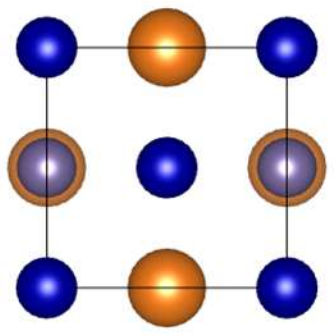 MgCoGe | 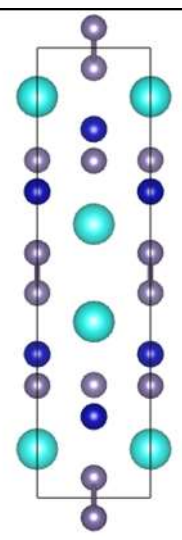 $YbCoGe_2$ |
| 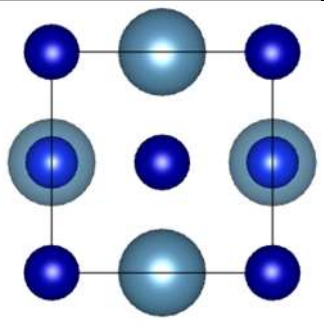 CaCoSi | 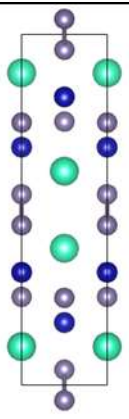 $LuCoGe_2$ | 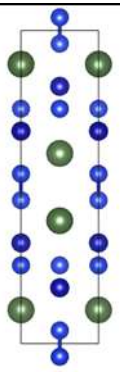 $ErCoSi_2$ |
| 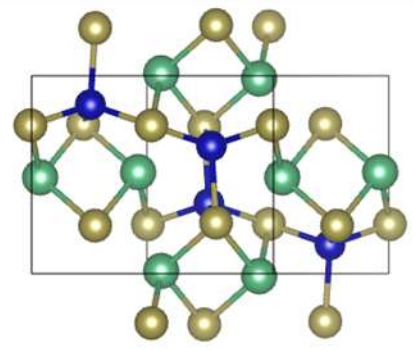 $NbCoTe_2$ | 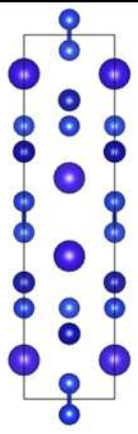 $DyCoSi_2$ | 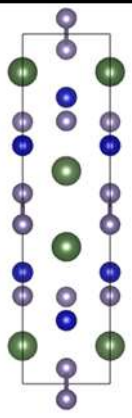 $ErCoGe_2$ |
| 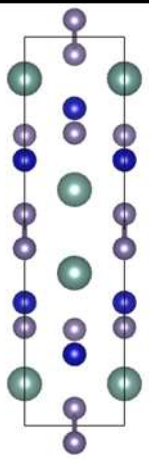 $YCoGe_2$ | 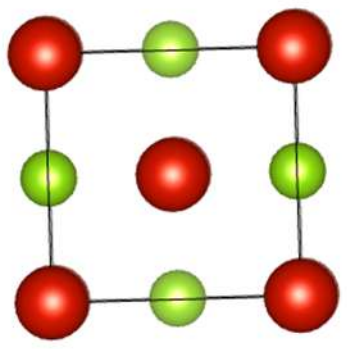 VSe | 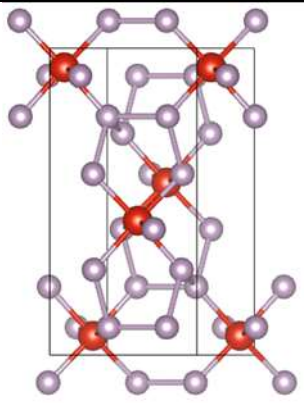 VP4 |
| 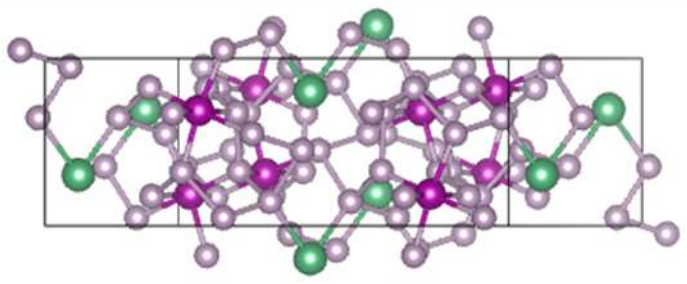 $Mn_2NbP_{12}$ | 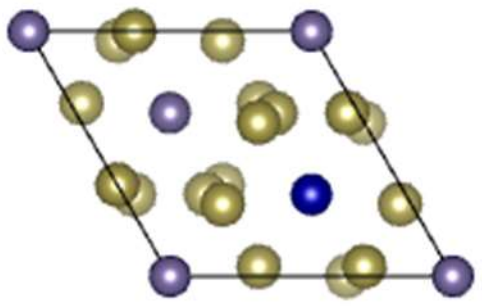 $CrGeTe_3$ | 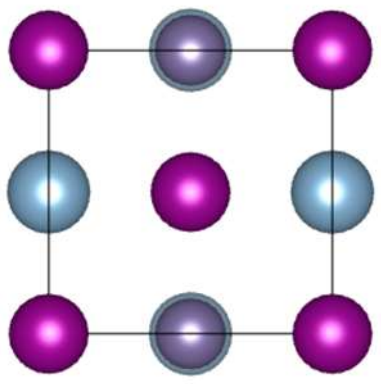 MnAlGe |
| 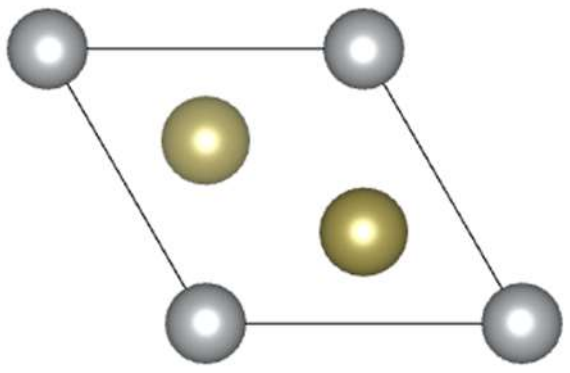 $NiTe_2$ | 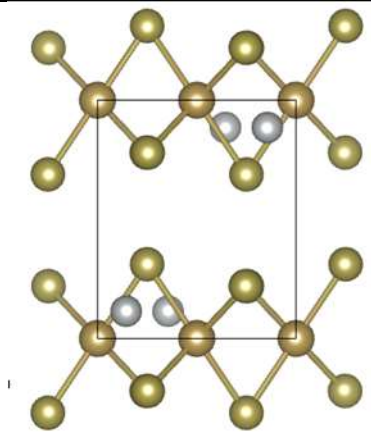 $TaNiTe_2$ | 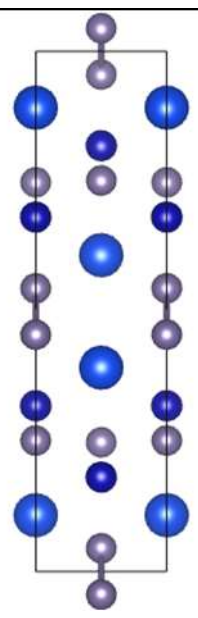 $HoCoGe_2$ |
| 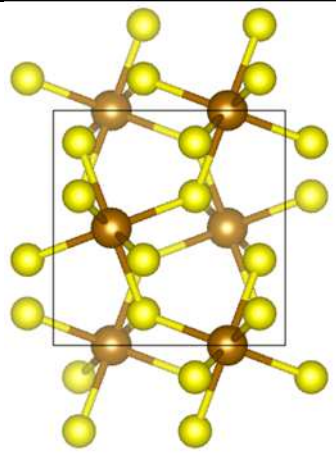 $FeS_2$ | 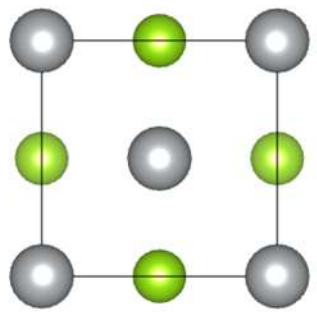 NiSe | 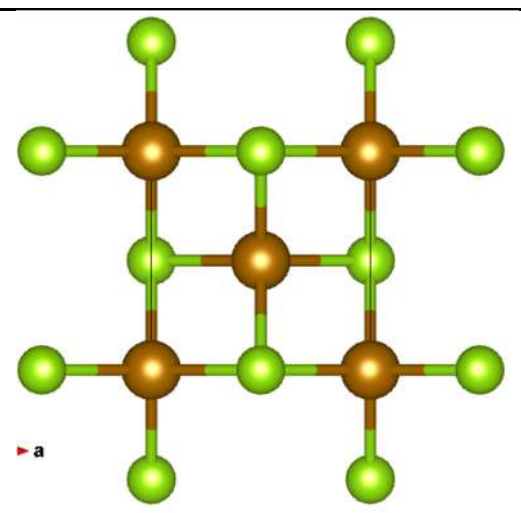 FeSe |
| 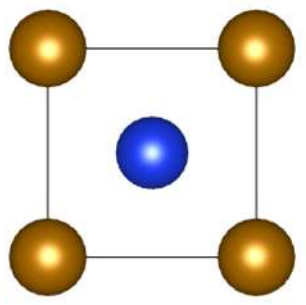 $FeSi_2$ | 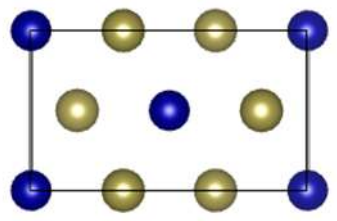 $Cr_3Te_4$ | 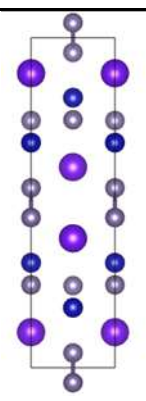 $TbCoGe_2$ |

***Ferromagnetic configuration***

| Name of Material | Total Energy | Magnetic Moment (Bohr Magneton) |
|---|---|---|
| $FeSi_2$ | -71.59911831 Ry | 0.3905 |
| $BaNiSn_3$ | -1330.66098584 Ry | 0.0019 |
| CaCoSi | -314.24413806 Ry | 0.0020 |
| CaNiGe | -345.05201142 Ry | 0.0028 |
| $Cr_3Te_4$ | -1175.04381764 Ry | 3.4888 |
| $CrGeTe_3$ | -1409.34667053 Ry | 3.1383 |
| $DyCoSi_2$ | -861.31394678 Ry | 0.0021 |
| $ErCoGe_2$ | -885.23826203 Ry | 0.0124 |
| $ErCoSi_2$ | -854.11554279 Ry | 0.0008 |
| $ErNiGe_2$ | -931.28659294 Ry | -0.0001 |
| $FeS_2$ | -387.08077362 Ry | 2.7773 |
| FeSe | -153.29297403 Ry | 0.2788 |
| $HoCoGe_2$ | -883.39161470 Ry | 0.0128 |
| $K_2CoSe_2$ | -471.72800815 Ry | 0.0123 |
| LiCoAs | -193.35070741 Ry | -0.0001 |
| $LuCoGe_2$ | -890.73846822 Ry | 0.0061 |
| MgCoGe | -425.94173271 Ry | 0.0214 |
| $Mn_2NbP_{12}$ | -3007.11369132 Ry | 0.2805 |
| MnAlGe | -449.82048306 Ry | 2.2230 |
| $NbCoTe_2$ | -900.46119587 Ry | -0.0138 |
| $NbNiTe_2$ | -946.43005926 Ry | -0.0008 |
| NiSe | -213.48233060 Ry | 0.1979 |
| $NiT_2$ | -118.60557998 Ry | 0.0079 |
| SiNiSb | -226.26359708 Ry | -0.0002 |
| $TaNiTe_2$ | -1042.93104763 Ry | 0.0021 |
| $TbCoGe_2$ | -890.52466967 Ry | 0.0118 |
| $VP_4$ | -397.90455303 Ry | 0.0190 |
| VSe | -97.13903643 Ry | 0.3111 |
| $YbCoGe_2$ | -847.63703330 Ry | 0.0000 |
| $YCoGe_2$ | -742.64032255 Ry | 0.0093 |

***Antiferromagnetic configuration***

| **Name of Material** | **Total Energy** | **Magnetic Moment (Bohr Magneton)** |
|---|---|---|
| $FeSi_2$ | | |
| $BaNiSn_3$ | -1330.66056278 Ry | 0.0019 |
| CaCoSi | -314.24323726 Ry | 0.0020 |
| CaNiGe | -343.38631830 Ry | -0.0000 |
| $Cr_3Te_4$ | -1174.94928674 Ry | 3.4217 |
| $CrGeTe_3$ | -1383.11881926 Ry | 3.0556 |
| $DyCoSi_2$ | -861.14865977 Ry | -0.0002 |
| $ErCoGe_2$ | -882.23881147 Ry | -0.0003 |
| $ErCoSi_2$ | -854.07589029 Ry | -0.0004 |
| $ErNiGe_2$ | -931.28989812 Ry | 0.0001 |
| $FeS_2$ | -387.09694778 Ry | 2.6797 |
| FeSe | -153.30166315 Ry | 1.6943 |
| $HoCoGe_2$ | -883.14802064 Ry | 0.0012 |
| $K_2CoSe_2$ | -471.75508283 Ry | 0.0000 |
| LiCoAs | -193.37869207 Ry | -0.0000 |
| $LuCoGe_2$ | -892.53830621 Ry | -0.0006 |
| MgCoGe | -425.94119405 Ry | -0.0003 |
| $Mn_2NbP_{12}$ | -3007.11334339 Ry | 0.3014 |
| MnAlGe | -450.30866286 Ry | 2.4075 |
| $NbCoTe_2$ | -900.45840849 Ry | 0.0017 |
| $NbNiTe_2$ | -943.62999605 Ry | 0.0024 |
| NiSe | -213.16810980 Ry | -0.0035 |
| $NiT_2$ | -118.57411740 Ry | -0.0000 |
| SiNiSb | -226.43808919 Ry | 0.0000 |
| $TaNiTe_2$ | -1042.93685587 Ry | 0.0015 |
| $TbCoGe_2$ | -890.54457574 Ry | -0.0006 |
| $VP_4$ | -397.90455303 Ry | 0.0624 |
| VSe | -97.13682196 Ry | 0.0537 |
| $YbCoGe_2$ | -847.92483268 Ry | -0.0000 |
| $YCoGe_2$ | -742.49069458 Ry | -0.0004 |

CaCoSi

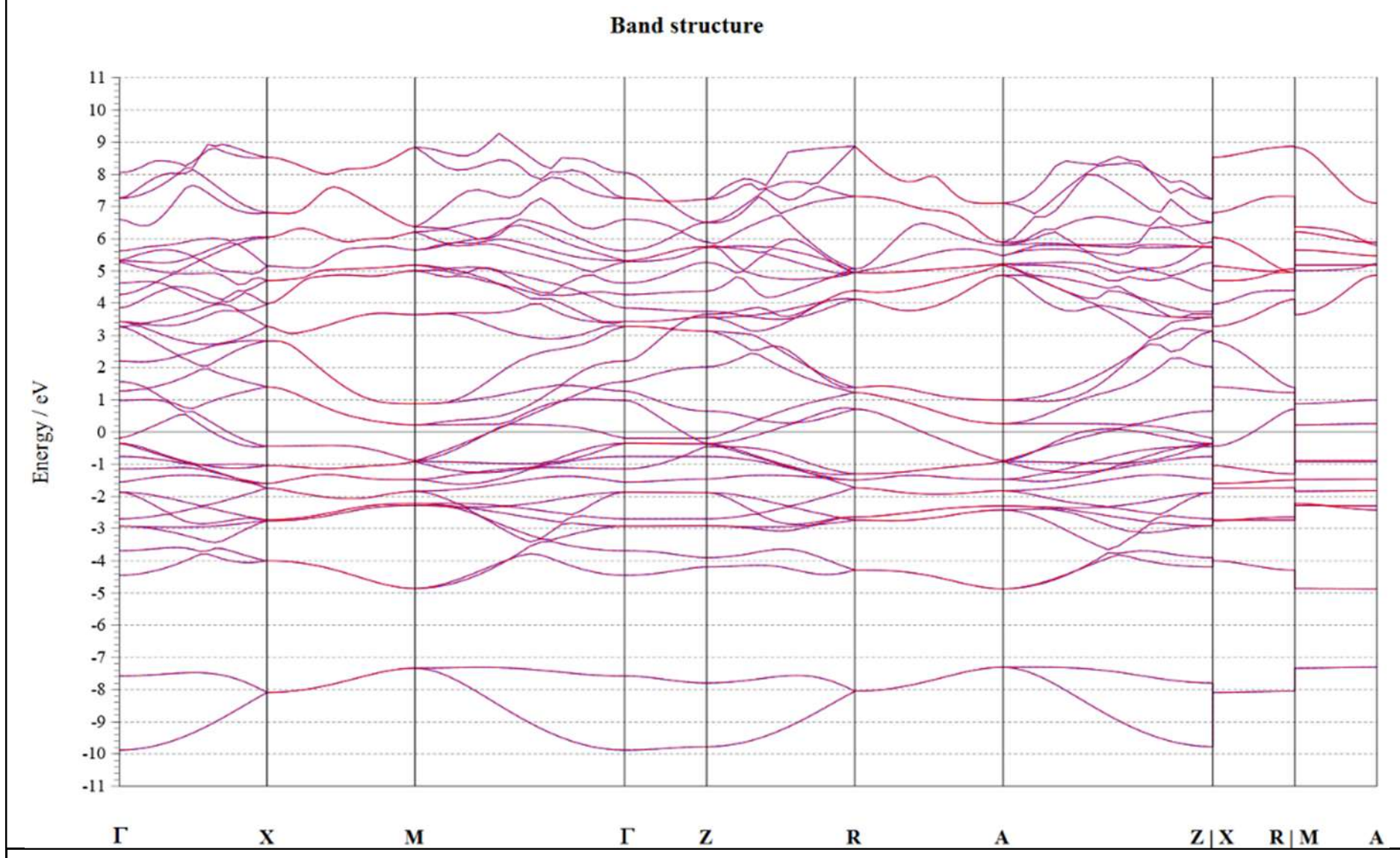

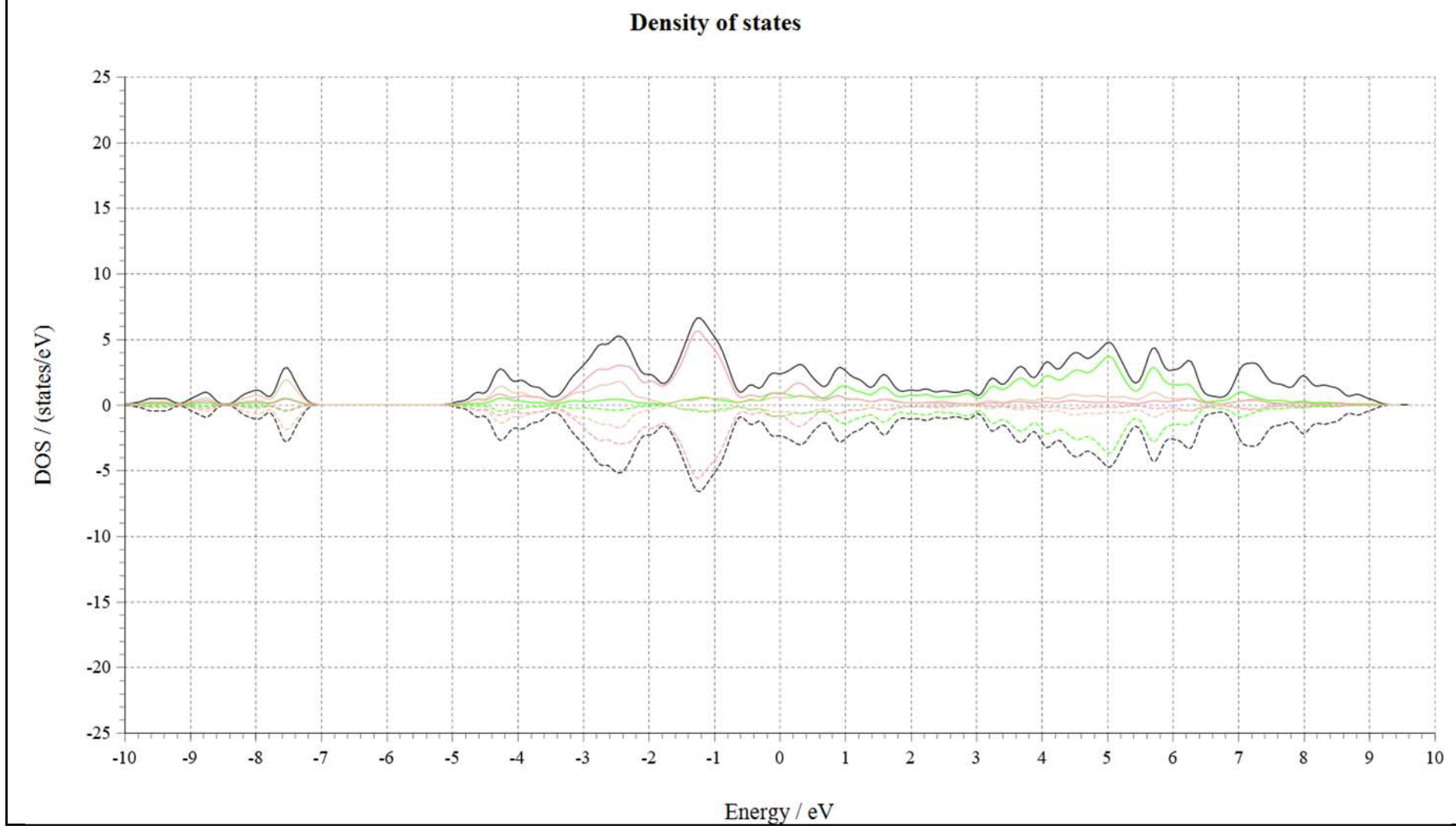

| CaNiGe |
| --- |
| 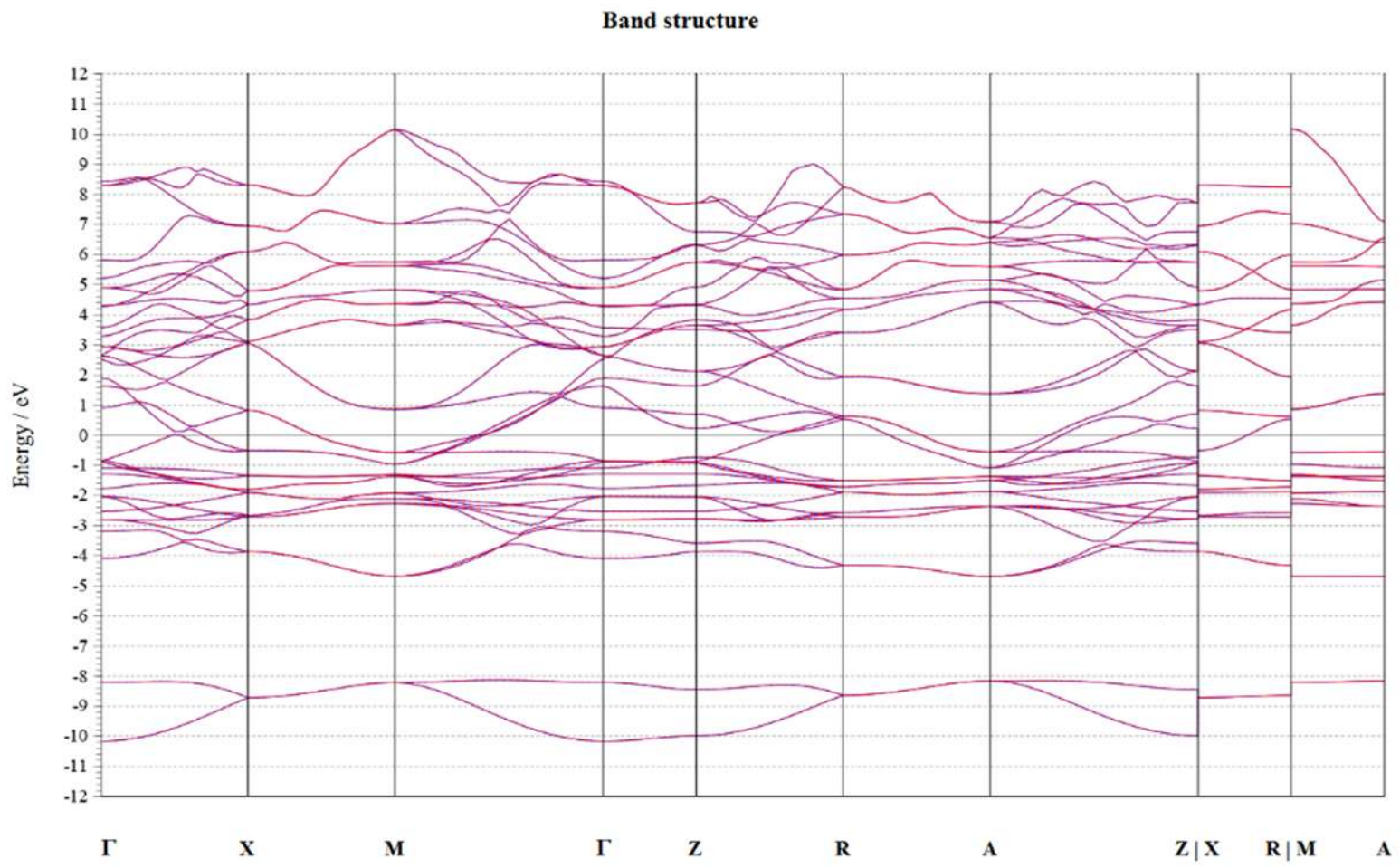Band structure<br>Energy / eV<br>Γ<br>X<br>M<br>Γ<br>Z<br>R<br>A<br>Z \| X<br>R \| M<br>A |
| 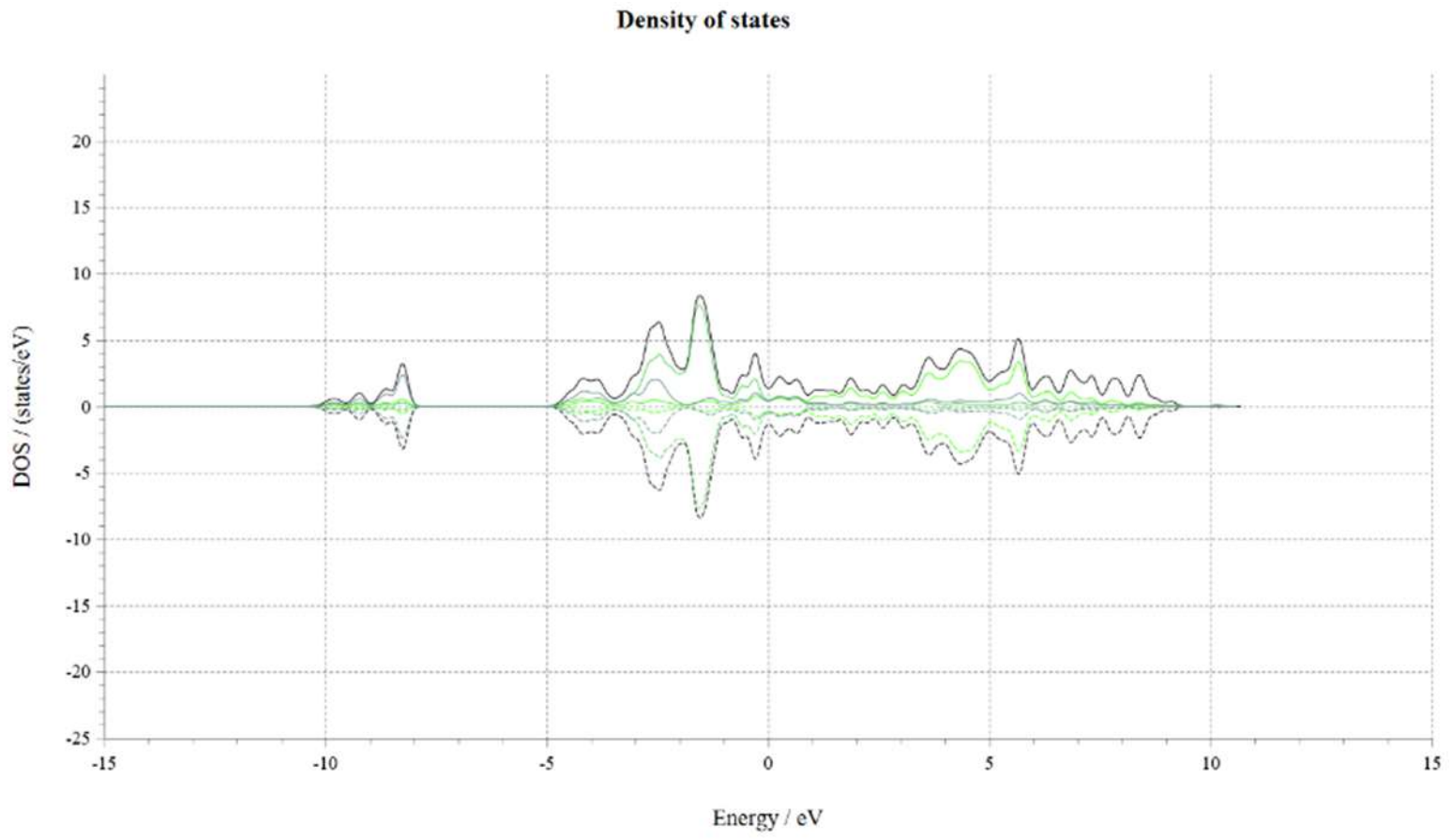Density of states<br>DOS / (states/eV)<br>Energy / eV |

| $Cr_3Te_4$ |
|---|

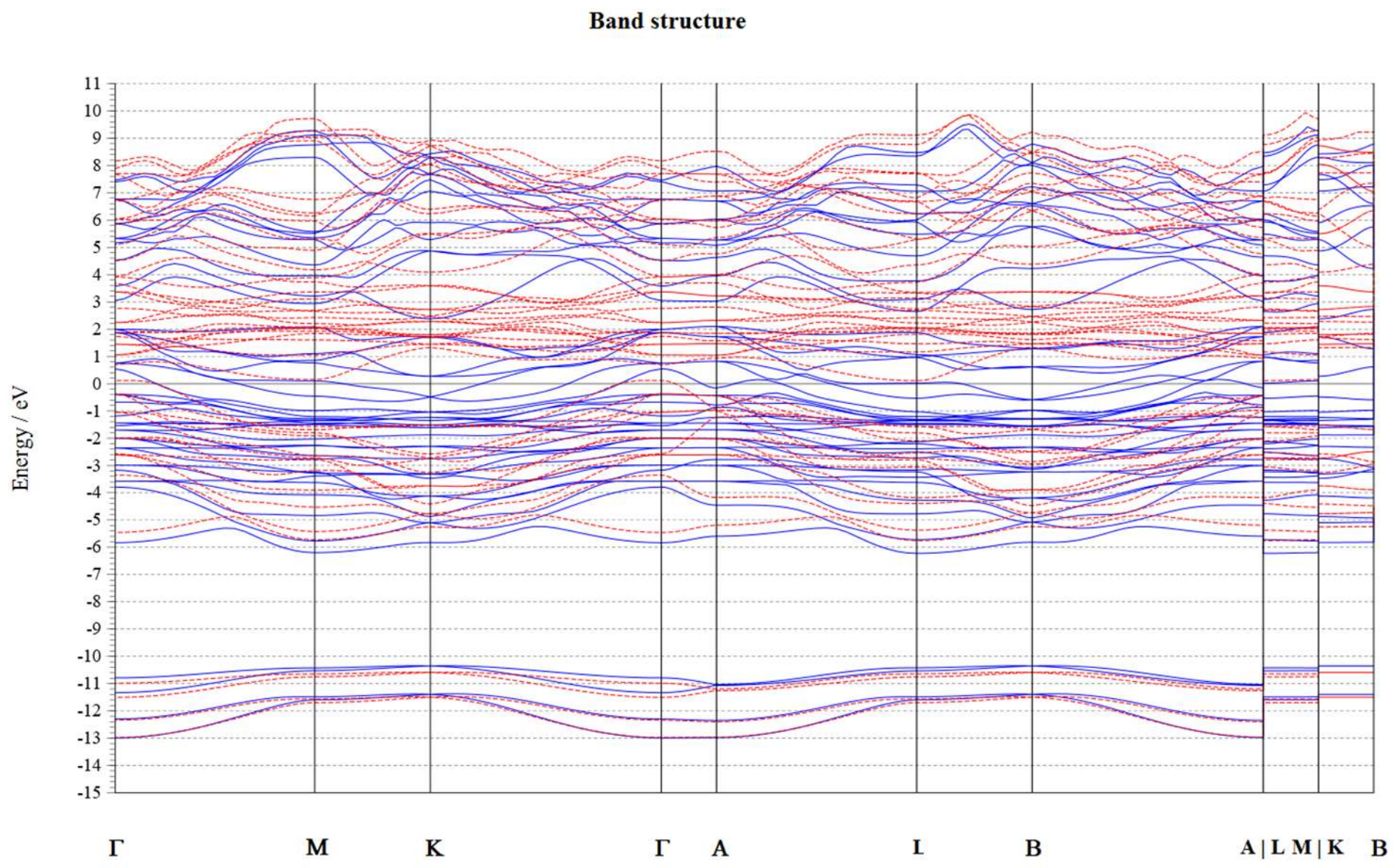

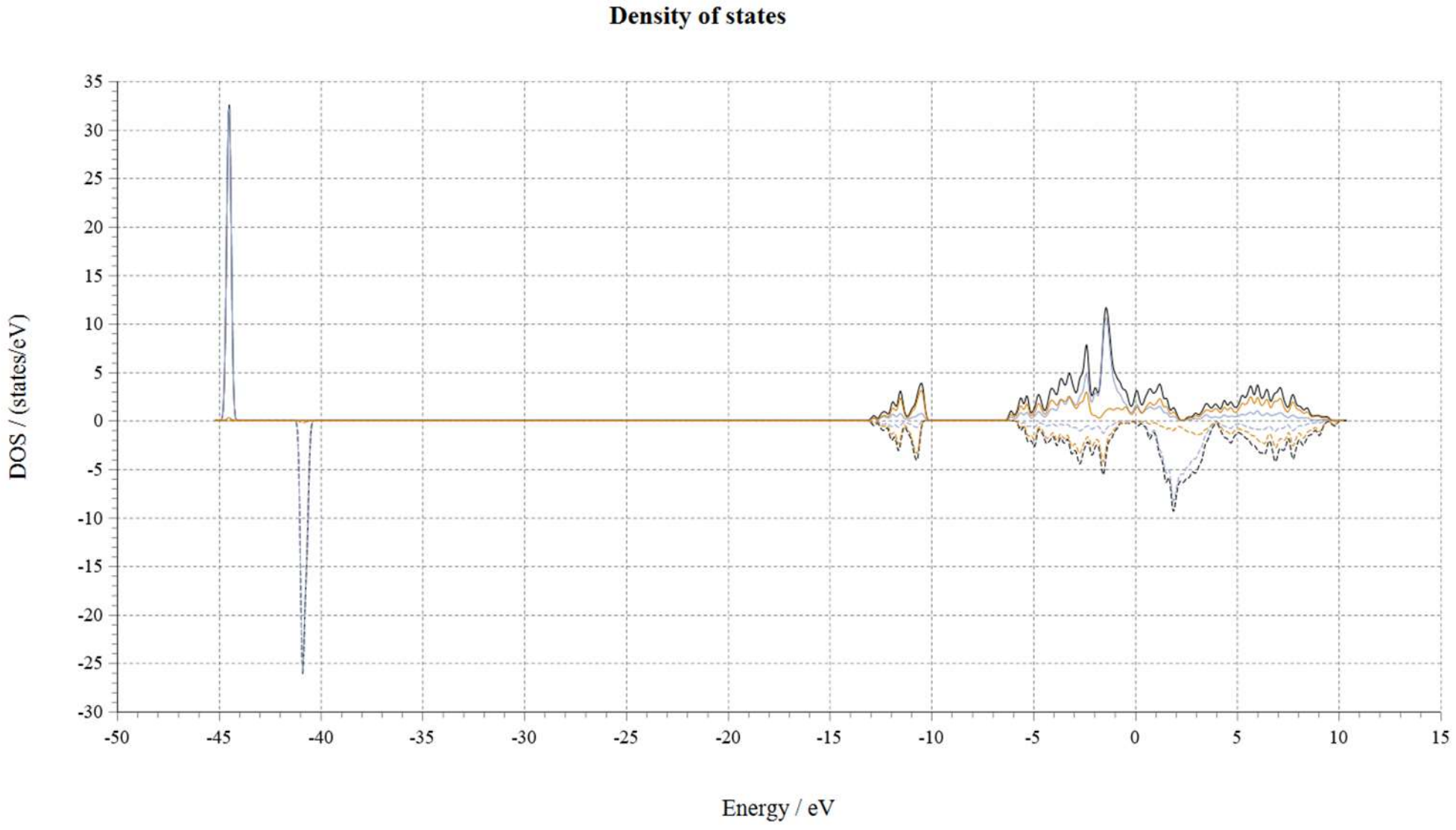

FeS$_2$

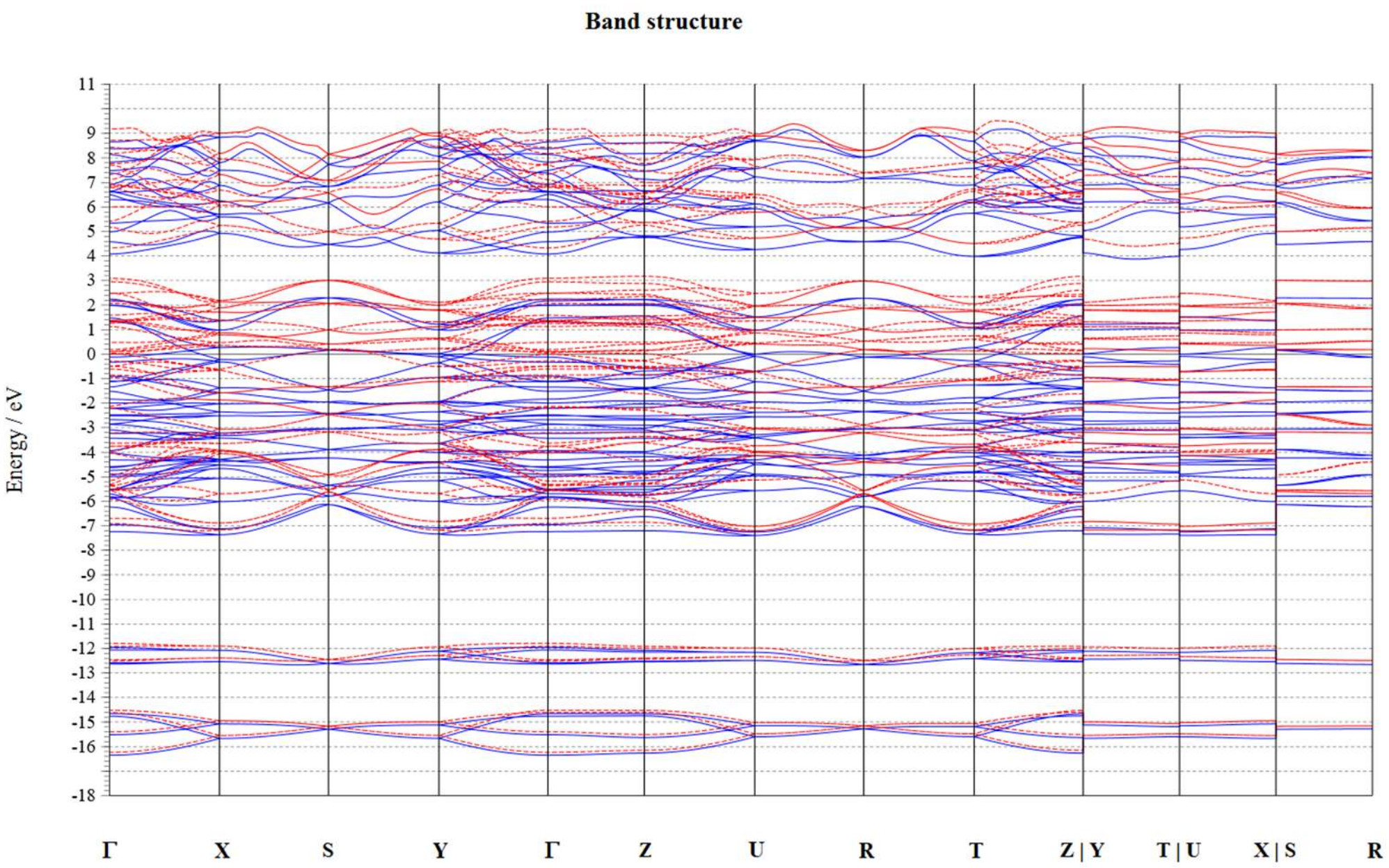

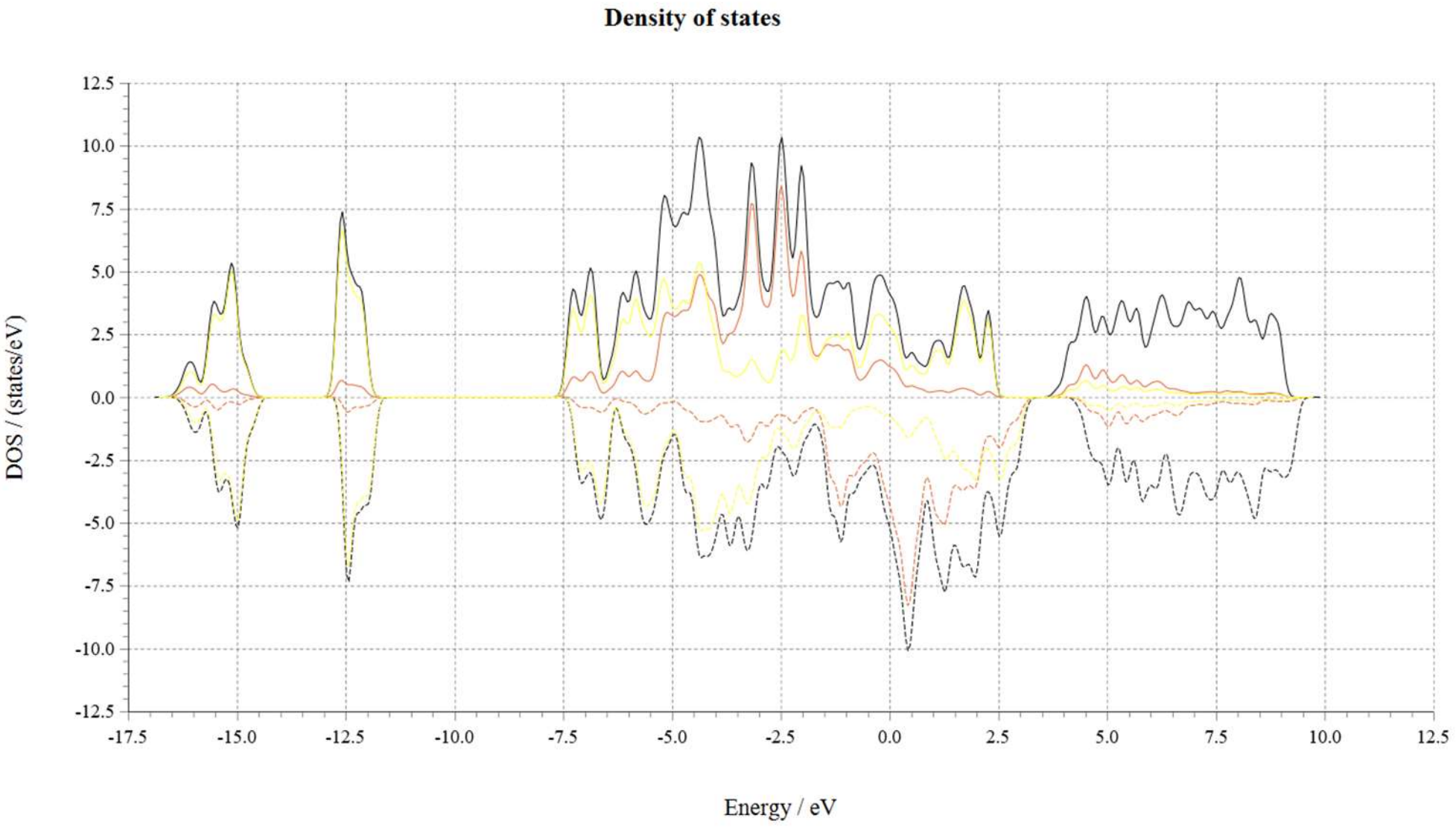

| FeSi$_2$ |
|---|

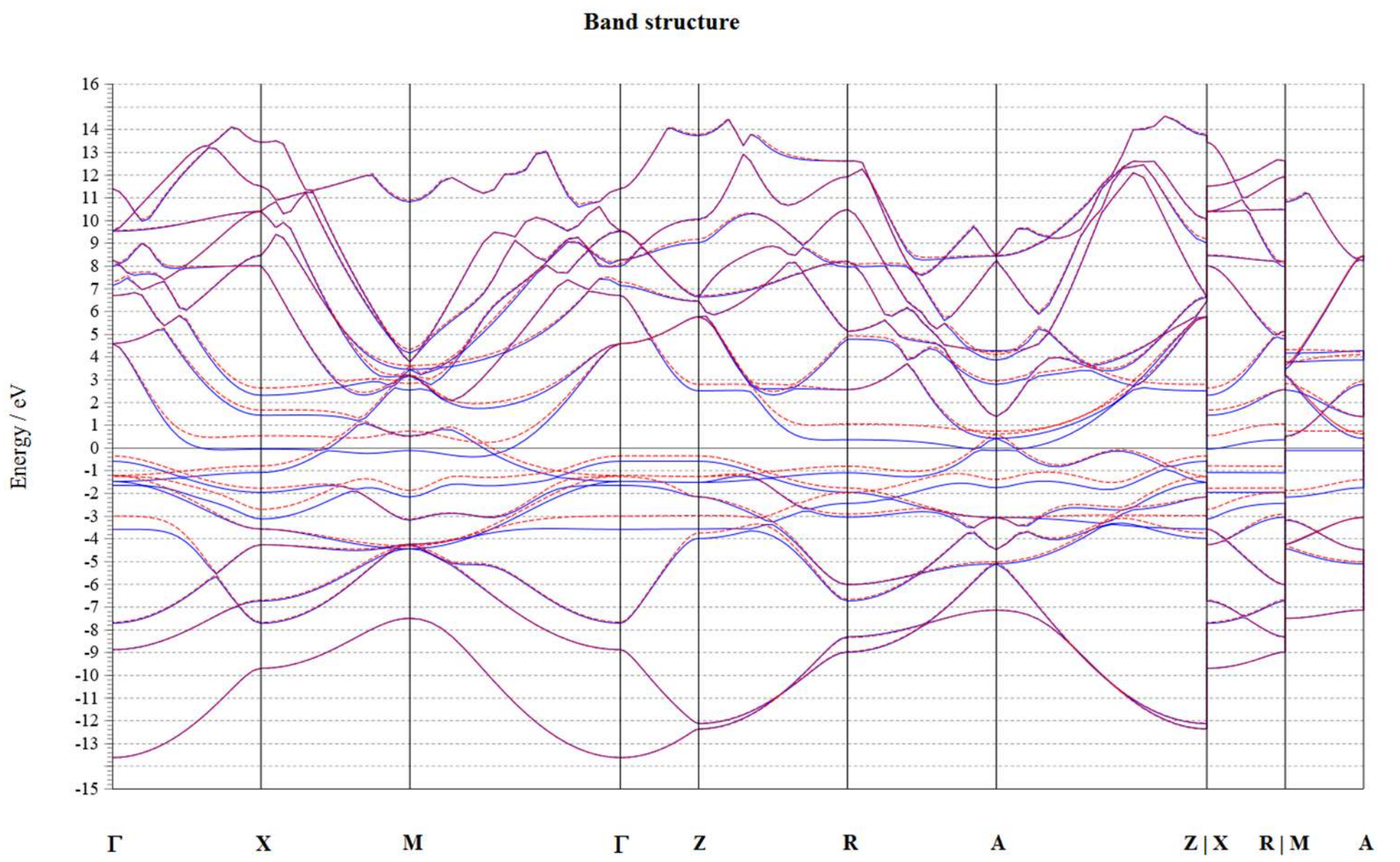

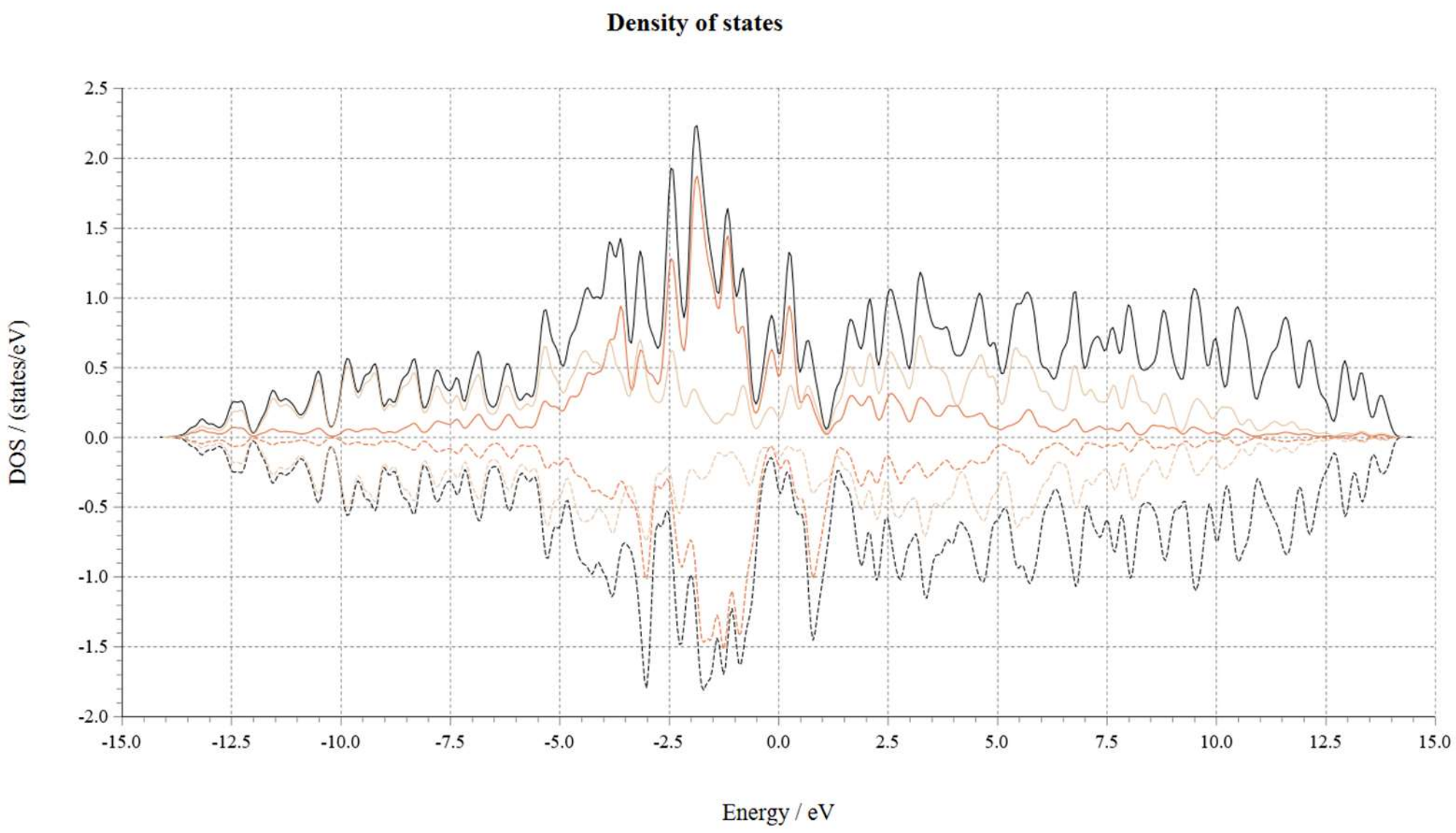
Density of states
DOS / (states/eV)
Energy / eV

| MnAlGe |
| --- |
| 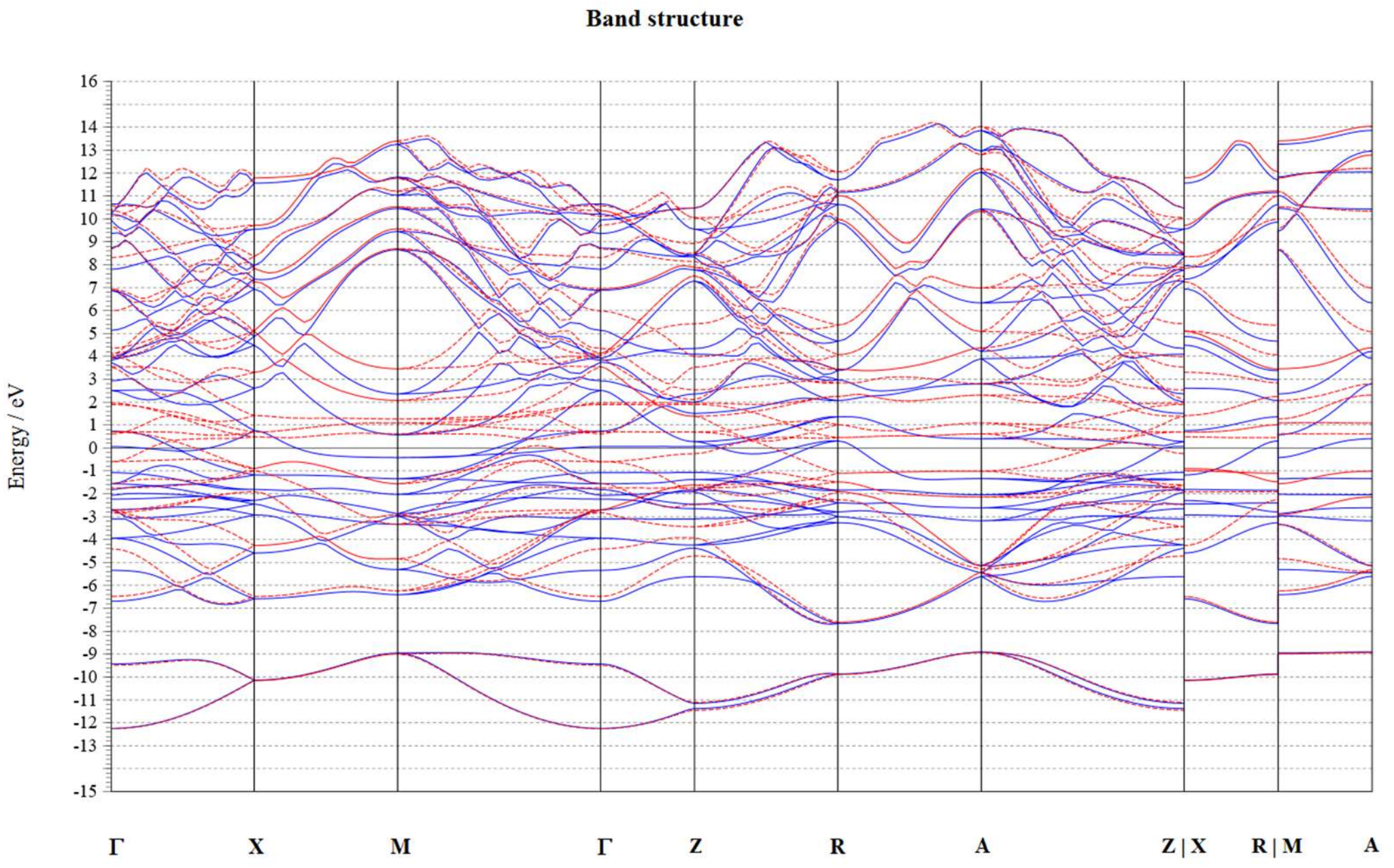 |
| 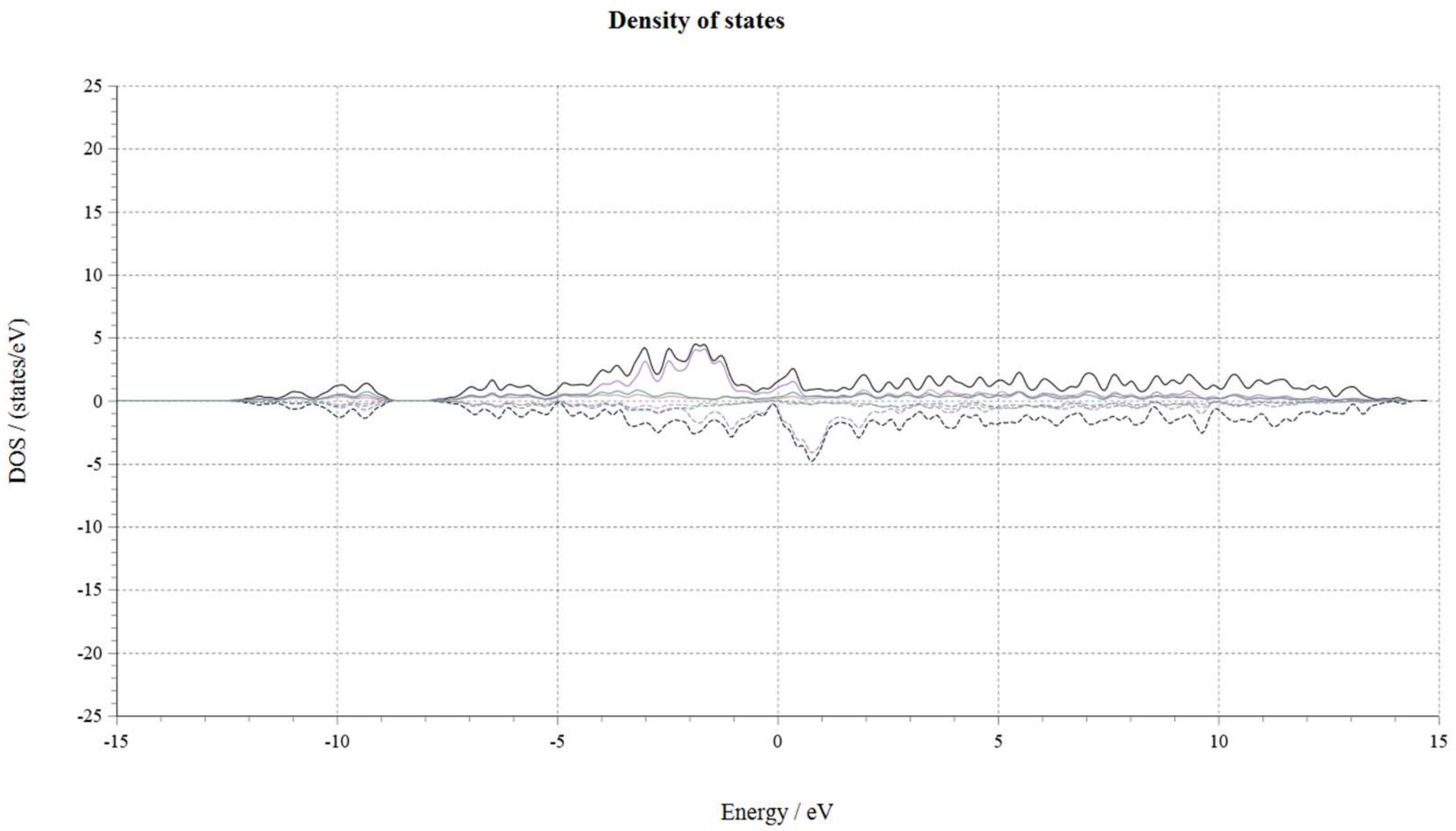 |

## 4. Conclusion

Many materials and devices that aid in the creation of tiny chips utilized in a variety of applications have been the subject of research over the past few decades [1-14]. Here two dimensional materials are first extracted from materials database. Then the materials were analyzed to find its properties.

Extracted parameters and properties can be used to simulate new memory devices and smaller chips.